\documentclass[letter]{aastex631}

\usepackage{graphicx}
\usepackage{graphbox}
\usepackage{hyperref}
\usepackage{amssymb,amsmath}
\usepackage{stmaryrd}
\usepackage{booktabs}
\usepackage{float}
\let\tablenum\relax
\usepackage[print-unity-mantissa=false, separate-uncertainty=true, multi-part-units=single]{siunitx}
\DeclareSIUnit\erg{erg}

\def\lea{\mathrel{<\kern-1.0em\lower0.9ex\hbox{$\sim$}}}
\def\gea{\mathrel{>\kern-1.0em\lower0.9ex\hbox{$\sim$}}}
\newcommand{\lta}{{\>\rlap{\raise2pt\hbox{$<$}}\lower3pt\hbox{$\sim$}\>}}
\newcommand{\gta}{{\>\rlap{\raise2pt\hbox{$>$}}\lower3pt\hbox{$\sim$}\>}}

\begin{document}
\title{Dataset of Classified Chandra Sources in Globular Clusters}

\shorttitle{GC CSC TD}

\author{Steven Chen}
\affiliation{Department of Physics, The George Washington University, 725 21st St. NW, Washington, DC 20052}
\email{schen70@gwmail.gwu.edu}

\author{Oleg Kargaltsev}
\affiliation{Department of Physics, The George Washington University, 725 21st St. NW, Washington, DC 20052}

\author{Hui Yang}
\affiliation{Department of Physics, The George Washington University, 725 21st St. NW, Washington, DC 20052}

\author{Jeremy Hare}
\affiliation{NASA Goddard Space Flight Center, Greenbelt, MD, 20771}
\affiliation{NASA Postdoctoral Program Fellow}

\begin{abstract}

We present a collection of classified X-ray sources in Globular Clusters (GCs) observed by the Chandra X-ray Observatory (CXO), including active binaries, cataclysmic variables, millisecond pulsars, and low-mass X-ray binaries. We cross-match the most accurate published positions from multiwavelength observations of these sources to the Chandra Source Catalog (CSC) Release 2.1, and the HST UV Globular Cluster Survey (HUGS) to extract their multiwavelength properties. The dataset can be accessed via an \href{https://home.gwu.edu/~kargaltsev/XCLASS_GC}{interactive website} and used as a training dataset for machine-learning classification of unidentified X-ray sources in GCs. 

\end{abstract}


\section{Introduction}

Globular clusters are ancient, massive, and dense star clusters orbiting the periphery of galaxies. About $150$ GCs are known in the Milky Way \citep{kharchenko_global_2013}. While densely populated with main sequence stars, GCs are also interesting as stellar graveyards containing thousands of compact objects, including white dwarfs, neutron stars (NSs), and black holes (BHs) in various types of systems. Over the past few decades, extensive multiwavelength observations of GCs have been performed revealing active binaries (ABs), cataclysmic variables (CVs), millisecond pulsars (MSPs), and low-mass X-ray binaries in quiescent (qLMXBs) and active states.  
However, 80\% of $\sim 2000$ CXO sources in all Galactic GCs remain unclassified. Many of these sources are likely to harbor compact objects, providing insights into stellar and binary evolution, as well as cluster dynamics. Efficient classification of CXO sources with supervised machine learning has been developed recently (e.g., \cite{yang_classifying_2022}), which requires a training dataset (TD) of reliably classified sources. It is difficult to classify X-ray sources solely from X-ray properties, especially since most of them are X-ray faint. Therefore, obtaining accurate multiwavelength properties, both for classified sources in the TD, and unclassified sources, is crucial. We present a dataset of reliably classified X-ray sources in Galactic GCs with their multiwavelength properties collected from CSC 2.1 \citep{2020AAS...23515405E} and HUGS \citep{nardiello_hubble_2018}. The dataset is available at \url{https://home.gwu.edu/~kargaltsev/XCLASS_GC} in an interactive environment.

\section{Methods}
\label{methods}

We compiled source positions, positional uncertainties (PUs; at 95\% confidence), and reliable classifications from 50+ publications\footnote{referenced in our dataset} on CXO sources in 23 GCs. We then sorted the X-ray sources into 5 main classes: ABs, CVs, MSPs, Spider MSPs, and qLMXBs. For the 19 clusters where HST counterparts to X-ray sources have been published, we also compiled HST counterpart coordinates, photometry, and instruments/filters used, when available. When multiple epochs of published data are available (e.g., for 47 Tuc, Omega Cen, and NGC 6397), we used the latest values. 

We found that the reduction and analysis of the CXO and HST data was quite heterogeneous among different publications. For uniformity, we cross-matched the published coordinates to the CSC and HUGS catalogs, which are both astrometrically tied to the Gaia catalogs (Gaia DR2 for CSC, Gaia DR1 for HUGS, \cite{brown_gaia_2018}) enabling reliable cross-matching between the two. 
 
We first cross-matched the published coordinates of CXO sources in GCs to CSC sources, using both PUs combined in quadrature. For cases where the published PUs were less than 0.5\arcsec or not reported, we used 0.5\arcsec. We found systematic (i.e., the same for all sources from a particular publication) astrometric offsets between the published and CSC source coordinates to be 0.1--0.3$\arcsec$ in most GCs. The offsets can be attributed to the differences between astrometric corrections performed in individual publications and the CSC. We dropped a few sources with CSC detection significance $<2$. In rare cases, a CSC source is cross-matched to multiple X-ray sources that are resolved in publications but not in CSC; we mark these sources with an ``Unresolved" flag. After cross-matching, we adopted the CSC properties, including coordinates, multi-band fluxes, variability indices, and hardness ratios. 

Similarly, we cross-matched the published coordinates of the HST counterparts to  HUGS, which provides magnitudes in 5 photometric bands (WFC3/UVIS F275W, F336W, F438W, and ACS/WFC F606W, F814W). As the density of HST sources in GCs is much higher than in the X-rays, and absolute HST astrometry in some (especially older) publications could be inaccurate, we used an iterative approach to find the correct HUGS counterpart. We first cross-matched published HST sources to the HUGS source with the smallest mean difference between their magnitudes (in the closest comparable bands) within a 1.5\arcsec radius, and plotted the separations. If a clear systematic offset between published coordinates and HUGS coordinates is seen, the median offset was applied as an astrometric correction to their published coordinates. Then, the closest HUGS source is taken for two cases: if its separation is $<0.1\arcsec$ ($<0.2\arcsec$) after the correction and the photometry differs by $<1$ mag in at least one (two) comparable band(s). This procedure was applied to 47 Tuc, where a known 1.3\arcsec astrometric offset \citep{clement_variable_2001} between published HST coordinates in \cite{albrow_frequency_2001} and true coordinates was confirmed; and NGC 6397, where a 0.3\arcsec offset was found. Other clusters either had good astrometry, or too few HST counterparts for determining a systematic shift, and a simple positional cross-match was used if it satisfied the above criteria. If no HST counterpart to an X-ray source was identified in publications, then only X-ray properties were used in the dataset. Since Omega Centauri was not included in HUGS, we used the Hubble Source Catalog v3 \footnote{HSC v3 astrometry is also referenced to Gaia DR1} \citep{lubow_hubble_2017}, taking photometry from the bands matching HUGS. 

We also cross-matched GC MSPs from the ATNF pulsar catalog \citep{manchester_australia_2005} to the CSC, when precise coordinates were available. We found that all, except one, cross-matches already had their X-ray counterparts reported in the publications we used. 
For detailed notes about the processing of data in Omega Cen and other clusters, and cross-matching to the ATNF catalog, see the auxiliary information on the website. 

\section{Summary}

We compiled from published works on CXO sources in GCs:

\begin{itemize}
    \item 268 CSC sources confidently classified as an AB, CV, MSP, qLMXB, or Spider MSP;
    \item 207 HST counterparts to CXO sources with reliable published classifications;
    \item 148 confident HUGS cross-matches to HST counterparts;\footnote{Other sources either did not match to a nearby HUGS source, or differences between published and HUGS magnitudes were too large.} 
    \item 46 MSPs cross-matched between CSC and ATNF.
\end{itemize}

We make our dataset publicly available online at \url{https://home.gwu.edu/~kargaltsev/XCLASS_GC}, using an interactive visualization tool CIDView\footnote{See https://github.com/ivv101/CIDview for details.}. Much of the functionality is similar to that described in  \cite{yang_visualizing_2021}, which focuses on classified CXO sources outside GCs. In brief, the tool allows users to manipulate the plot by selecting different classes, source features, and individual sources or a subset of sources. A table of selected sources with detailed information can be displayed, together with their multiwavelength images via \href{https://aladin.cds.unistra.fr/AladinLite/}{AladinLite} and \href{https://sky.esa.int/esasky/}{ESASky} widgets. For detailed instructions, see the \textit{readme} on the website.

\begin{figure*}
    \centering
    \includegraphics[width=\textwidth]{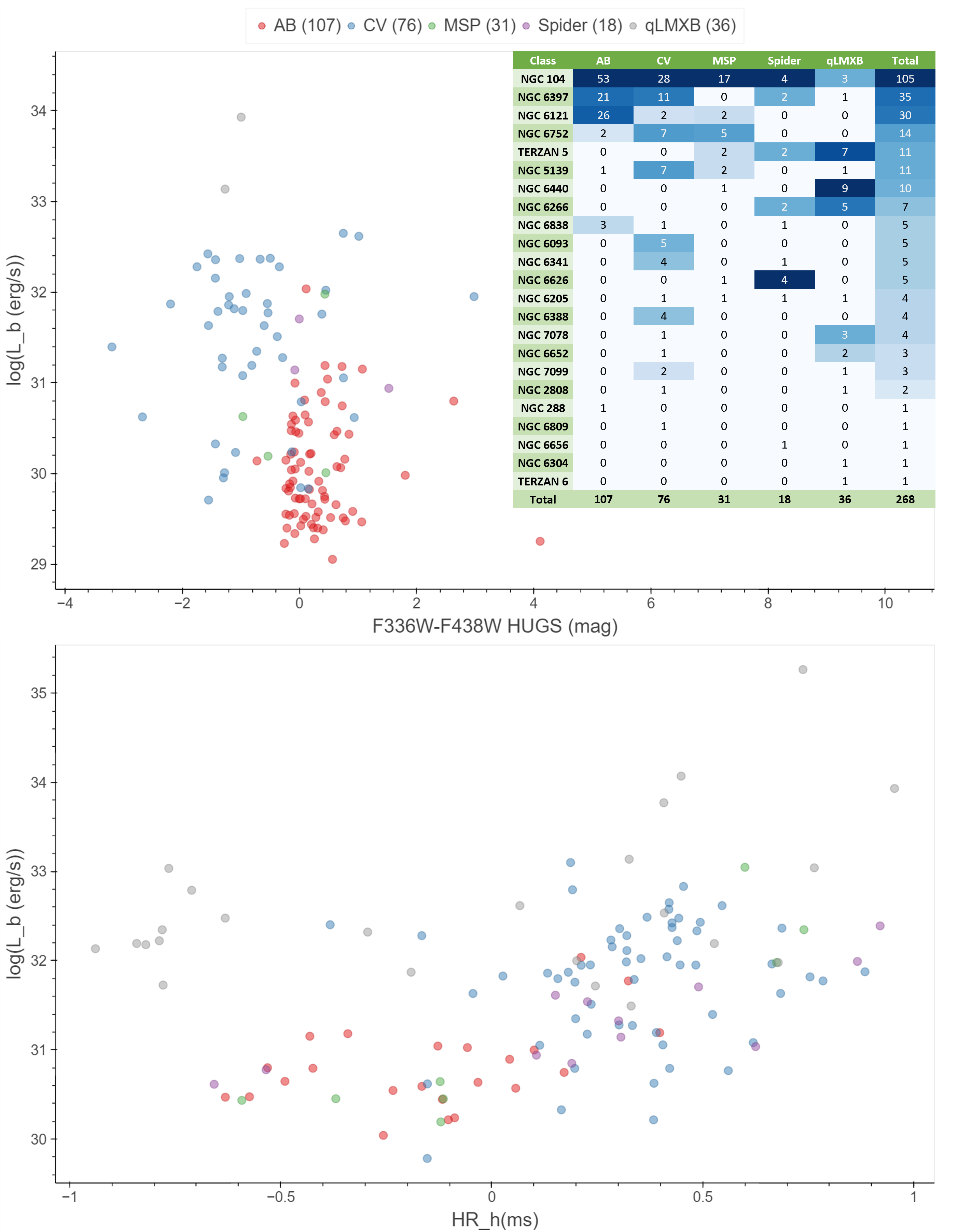}
    \caption{Screenshot of the visualization tool, showing two plots with different features selected. The inset table shows a summary of classified sources per cluster, with color shading corresponding to number of sources per cluster, normalized by class.}
    \label{fig:td_website}
\end{figure*}

In Figure \ref{fig:td_website}, we show example plots of X-ray broadband luminosity vs.\ F336W-F438W color, and hardness ratio HR${}_{\mathrm{h(ms)}}$. The inset summarizes the number of sources in different classes per cluster. Although the features available in the tool are limited to CXO fluxes, luminosities, variability indices, hardness ratios, and selected absolute magnitudes and colors (HUGS bands, and F555W, F625W, F658N), the full dataset, with all features, can be downloaded.

Support for this work was provided through HST-AR-16620.001 grant from
the STScI under NASA contract NAS5-26555.

\newpage

\bibliographystyle{aasjournal}
\bibliography{references.bib}

\end{document}